\documentclass[twocolumn]{jpsj3} 
\usepackage{bm}

\title{Microscopic Study of Electronic and Magnetic Properties for Ir Oxide}

\author{Tomonori \textsc{Shirakawa}\thanks{E-mail: t-shirakawa@riken.jp}, Hiroshi \textsc{Watanabe}, and Seiji \textsc{Yunoki}}

\inst{Computational Condensed Matter Physics Laboratory, RIKEN ASI, Wako, Saitama 351-0198, Japan, \\
CREST, Japan Science and Technology Agency (JST), Kawaguchi, Saitama 332-0012, Japan, and \\
Computational Materials Science Research Team, RIKEN AICS, Kobe, Hyogo 650-0047, Japan}

\abst{
The exact diagonalization and the variational cluster approximation (VCA) 
are used to study the nature of a novel Mott insulator induced by a strong spin-orbit coupling 
for a two-dimensional three-band Hubbard model consisting of the $t_{2g}$ manifold of $5d$ orbitals. 
To characterize the ground state, we introduce a local Kramer's doublet 
which can represent a state with effective angular momentum $J_{\rm eff}=\left|{\bm S}-{\bm L}\right|=1/2$ 
as well as spin $S=1/2$. 
Our systematic study of the pseudo-spin structure factor defined by the Kramer's doublet shows that the 
$J_{\rm eff}=1/2$ Mott insulator is smoothly connected to the $S=1/2$ Mott insulator.  
Using the Kramer's doublet as a variational state for the VCA, we examine the one-particle excitations 
for the Mott insulating phase.  These results are compared with recent experiments 
on Sr$_2$IrO$_4$. 
 }

\kword{5d electrons, spin-orbit coupling, Mott insulator, three-band Hubbard model, 
variational cluster approximation}

\begin{document}
\maketitle
\section{Introduction}
5$d$ transition metal oxides have attracted much attention because of 
their unique properties caused by a strong spin-orbit 
coupling (SOC) $\lambda$ for 5$d$ transition elements. 
One of such materials is Sr$_2$IrO$_4$ 
in a layered perovskite structure of K$_2$NiF$_4$ type.~\cite{randall,crawford} 
While Sr$_2$IrO$_4$ was first synthesized more than fifty years ago,~\cite{randall} 
it is only in 90's that its electronic properties has been studied systematically 
as an analogous system to high-T$_c$ cuprate superconductors. 
An inelastic neutron scattering experiment has found that the ground state of 
Sr$_2$IrO$_4$ is antiferromagnetically ordered with weak ferromagnetic moment, 
similar to that for the parent compounds of cuprates.~\cite{crawford} 
The early studies then concluded that the ground state of Sr$_2$IrO$_4$ was a spin $S=1/2$ 
Mott insulator with unpaired electrons occupying a half-filled $d_{xy}$ band, which 
is split off upward compared to $d_{yz}$ and $d_{zx}$ orbitals due to the crystalline electrostatic 
field with elongation of the Ir-O bond along $z$ direction.~\cite{cao} 
However, very recently, x-ray scattering experiments have revealed 
that the ground state is instead close to a $J_{\rm eff}=1/2$ Mott insulator.~\cite{kim1} 
Here, $J_{\rm eff}=\left|{\bm S}-{\bm L}\right|$ is an effective total angular 
momentum defined in the $t_{2g}$ manifold with the orbital angular momentum ${\bm L}$. 
Note also that the $J_{\rm eff}=1/2$ state corresponds to 
the ground state in the atomic limit with large $\lambda$. 


Motivated by these experiments, we shall study theoretically the electronic and magnetic properties 
of Sr$_2$IrO$_4$ using a three-band Hubbard model with the SOC.~\cite{watanabe} 
First, we introduce a local Kramer's doublet which can represent a $S=1/2$ state as well as 
a $J_{\rm eff}=1/2$ state.~\cite{jackeli} 
Employing the exact diagonalization method, we show that the local Kramer's doublet 
can describe smoothly both the $S=1/2$ and $J_{\rm eff}=1/2$ Mott insulators with varying $\lambda$.  
This strongly indicates that there exists no apparent symmetry change between these two extreme states.  
We then employ the variational cluster approximation~\cite{vca} (VCA) method
based on the self-energy functional theory~\cite{sft} (SFT) to examine the one-particle 
excitations in the Mott insulating phase.  
We find that, for a realistic set of model parameters for Sr$_2$IrO$_4$, most of the unoccupied state, 
i.e., the upper Hubbard band, can be well describe by the $J_{\rm eff}=1/2$ state. 
This result is in good qualitative agreement with the recent experimental observation.~\cite{kim1} 
We also discuss the effects of SOC and local 
Coulomb interactions on the one-particle excitations. 

The paper is organized as follows. After describing the three-band Hubbard model in Sec.~\ref{sec:model}, 
the local Kramer's doublet is introduced in Sec. \ref{sec:ed} and compared with the numerically exact ground state 
of the model.  Using VCA based on SFT, the one-particle excitations are studied  in Sec.~\ref{sec:vca}. 
The paper is concluded in Sec.~\ref{sec:conc}. 

\section{\label{sec:model}Model}
Since the crystalline electrostatic field is much larger than the SOC and the local Coulomb 
interactions,~\cite{jin} the local electronic configuration of Ir$^{4+}$ ion in Sr$_2$IrO$_4$ is the 
low-spin state of $(t_{2g})^5$.  Therefore, we consider the following 
effective three-band Hubbard model, consisting of three $d$ orbitals ($d_{xy}$, $d_{yz}$, and $d_{zx}$), 
on the square lattice 
\begin{eqnarray}
H &= &H_{\rm kin} + H_{\rm so} + H_{\rm int} \\
H_{\rm kin} &=& \sum_{{\bm k}\alpha \sigma} \epsilon_{\bm k}^{\alpha} 
c_{{\bm k}\alpha \sigma}^{\dagger} c_{{\bm k}\alpha \sigma} \\
H_{\rm so} &=& \lambda \sum_{\bm r} {\bm \ell}_{\bm r} \cdot {\bm s}_{\bm r} \\
H_{\rm int} &=& U \sum_{{\bm r}\alpha} n_{{\bm r}\alpha \uparrow} n_{{\bm r}\alpha \downarrow} 
+ \frac{U^{\prime}}{2} \sum_{{\bm r}\sigma} \sum_{\alpha \neq \beta}
n_{{\bm r}\alpha \sigma} n_{{\bm r}\beta \bar{\sigma}} \nonumber \\
&{}& + \frac{1}{2} (U^{\prime} - J) \sum_{{\bm r}\sigma} \sum_{\alpha \neq \beta} 
n_{{\bm r}\alpha \sigma} n_{{\bm r}\beta \sigma} \nonumber \\
&{}& - J \sum_{\bm r} \sum_{\alpha \neq \beta}
c_{{\bm r}\alpha \uparrow}^{\dagger}c_{{\bm r}\alpha \downarrow}
c_{{\bm r}\beta \downarrow}^{\dagger} c_{{\bm r}\beta \uparrow} \nonumber \\
&{}& + J^{\prime} \sum_{\bm r} \sum_{\alpha \neq \beta} 
c_{{\bm r}\alpha \uparrow}^{\dagger} c_{{\bm r}\alpha \downarrow}^{\dagger} 
c_{{\bm r}\beta \downarrow} c_{{\bm r}\beta \uparrow}.
\end{eqnarray}
Here, $c_{{\bm r}\alpha \sigma}$ ($c_{{\bm r}\alpha \sigma}^{\dagger}$) is the 
annihilation (creation) operator of an electron with 
spin $\sigma$ ($\sigma=\uparrow,\downarrow$) and orbital $\alpha$ ($\alpha = xy$, $yz$, and $zx$) 
at site ${\bm r}$.  $c_{{\bm k}\alpha \sigma}$ is the Fourier transform of $c_{{\bm r}\alpha \sigma}$. 
$\epsilon_{\bm k}^{\alpha}$ is the dispersion of orbital $\alpha$ 
\begin{eqnarray}
&{}& \epsilon_{\bm k}^{xy} = -2 t_1 (\cos k_x + \cos k_y) - t_2 \cos k_x \cos k_y \nonumber \\
&{}& \quad \quad \quad - 2 t_3 (\cos 2 k_x + \cos 2 k_y) + \Delta, \\
&{}& \epsilon_{\bm k}^{yz} = -2 t_4 \cos k_y - 2 t_5 \cos k_x, \\
&{}& \epsilon_{\bm k}^{zx} = - 2 t_4 \cos k_x - 2 t_5 \cos k_y, 
\end{eqnarray}
where $t_1$, $t_2$, and $t_3$ correspond to the hopping integrals for $d_{xy}$ orbitals 
located at the nearest, next nearest, and third nearest neighbor sites, 
respectively. $t_4$ and $t_5$ are the nearest neighbor hopping integrals for  
$d_{yz}$ ($d_{zx}$) orbital in $y$ ($x$) and $x$ ($y$) directions, respectively.  
$\Delta$ is an energy level difference between $d_{xy}$ orbital and the other orbitals ($d_{yz}$ and $d_{zx}$), 
which is naturally expected due to the large crystalline electrostatic field inducing tetragonal splitting. 
$H_{\rm so}$ is the SOC term represented in the following matrix form 
\begin{eqnarray}
2 {\bm \ell}_{\bm r} \cdot {\bm s}_{\bm r} = 
\sum_{\sigma} \left( c_{{\bm r}xy \bar{\sigma}}^{\dagger}, 
c_{{\bm r}yz \sigma}^{\dagger}, 
c_{{\bm r}zx \sigma}^{\dagger} \right) \times \nonumber \\
\left(
\begin{array}{ccc}
0 & -s & -i \\
-s & 0 & is \\
i & -is & 0 \\
\end{array}
\right)\left( 
\begin{array}{c}
c_{{\bm r}xy \bar{\sigma}} \\
c_{{\bm r}yz \sigma} \\
c_{{\bm r}zx \sigma} \\
\end{array}
\right),
\end{eqnarray}
where $s = +1$ ($-1$) for $\sigma = \uparrow$ ($\downarrow$), and $\bar{\sigma}$ indicates the 
opposite spin of $\sigma$. Finally, we introduce, for the local 
Coulomb interactions, the intra-orbital ($U$) and the inter-orbital ($U^{\prime}$) Coulomb 
interactions, the Hund's coupling $J$, and the pair-hopping $J^{\prime}$, with 
$U=U^{\prime}+2J$ and $J=J^{\prime}$~\cite{kanamori}. The number of electrons is  
set to be 5 per site. 

\begin{table}[tb]
\caption{Two sets of parameters used.}
\label{t1}
\begin{tabular}{ccc}
\hline
Parameters & Simplified model & Sr$_2$IrO$_4$ \\
\hline
$t_1$ & 0.36 eV & 0.36 eV \\
$t_2$ & 0.00 eV & 0.18 eV \\
$t_3$ & 0.00 eV & 0.09 eV \\
$t_4$ & 0.36 eV & 0.37 eV \\
$t_5$ & 0.00 eV & 0.06 eV \\
$\Delta$ & 0.36 eV & $-$0.36 eV \\
$\lambda$ & --- & 0.37 eV \\
\hline
\end{tabular}
\end{table}
As shown in Table~\ref{t1}, we use two sets of parameters indicated by 
"Simplified model'' and "Sr$_2$IrO$_4$". The parameter set "Sr$_2$IrO$_4$" is a 
realistic set of parameters determined by fitting the band dispersion obtained from the first 
principles calculations based on density functional theory.~\cite{jin} 

\section{\label{sec:ed}Exact diagonalization study}
\begin{figure}[tb]
\begin{center}
\includegraphics[width=0.9\hsize]{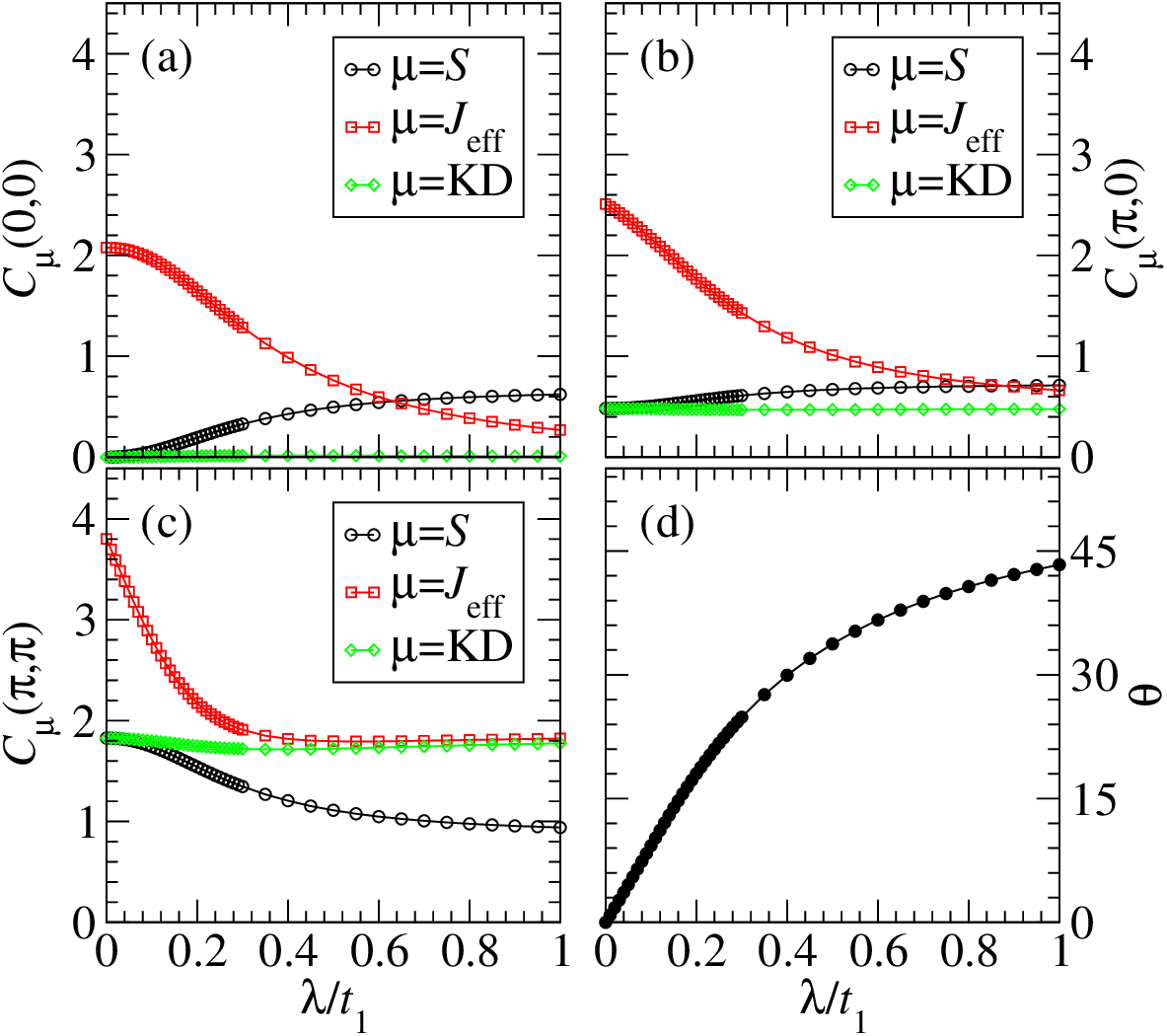}
\caption{(Color online) 
Structure factors $C_{\mu} ({\bm q})$ (for definition, see in the text) for (a) ${\bm q}=(0,0)$, (b) ${\bm q}=(\pi,0)$, 
and (c) ${\bm q}=(\pi,\pi)$. 
(d) The optimal value of $\theta$ for the Kramer's doublet [Eq.~(\ref{eq:kd})]. The model parameters used 
are $U/t_1 = 8$, $J/U = 0.15$, $\Delta / t_1 = 1$, and the simplified hopping listed in 
Table~\ref{t1}. 
}
\label{fig1}
\end{center}
\end{figure}
First, we have obtained the phase diagram for the simplified model with $U/t_1 = 8$ and $J/U = 0.15$ 
using exact diagonalization technique for a 4-site cluster. 
Let us briefly summarize the phase diagram of $\Delta$ vs. $\lambda$.~\cite{shirakawa} 
In region (I) $\Delta/t_1 \lesssim -0.3$ and $\lambda/t_1 \lesssim 0.15$, 
$d_{yz}$ and $d_{zx}$ orbitals are magnetically active, whereas most of $d_{xy}$ orbital is fully occupied. 
The degeneracy of $d_{yz}$ and $d_{zx}$ orbital leads to a ferromagnetic state with 
antiferro orbital ordering, which is expected by the second-order perturbation theory 
from the strong coupling limit in the presence of Hund's coupling.~\cite{oafsf} 
In region (II) $-0.3 \lesssim \Delta / t_1 \lesssim 0.4$ and $\lambda / t_1 \lesssim 0.2 $, 
the hole density is $n_{xy} : n_{yz}+n_{zx} = 1/2:1/2$ for $\lambda = 0$, and stripe-like spin correlations, 
characterized by ${\bm q}=(\pi,0)$ and $(0,\pi)$, become dominant. 
Finally, in region (III) $\Delta /t_1 \sim 1$ or $\lambda /t_1 \gtrsim 0.2$, 
an antiferromagnetic insulating phase appears, including both $S=1/2$ 
and $J_{\rm eff}=1/2$ Mott insulators in the limit of $\lambda=0$ and $\infty$, respectively. 

Let us now discuss closely the ground state properties in region (III). 
We first introduce the following local Kramer's doublet 
\begin{eqnarray}
a_{{\bm r}\theta \sigma}^{\dagger} = s \cos \theta c_{{\bm r}xy\sigma}^{\dagger} + \frac{\sin \theta }{\sqrt{2}}
\left( c_{{\bm r}yz\bar{\sigma}}^{\dagger} +  i s c_{{\bm r}zx\bar{\sigma}}^{\dagger} \right), 
\label{eq:kd}
\end{eqnarray}
which includes the two limiting states, namely, $S=1/2$ ($d_{xy}$) for $\theta = 0$ and $J_{\rm eff}=1/2$ 
for $\theta = \arctan \sqrt{2} \sim 54.74^{\circ}$. 
In the atomic limit, the eigenstate of the highest level of $H_{\rm so}$ 
in the presence of $\Delta$ is generally expressed by this state.~\cite{jackeli} 
Here, we shall optimize $\theta$ by 
minimizing the hole density $n_{\theta \sigma} = \left< \psi_0 \right| a_{{\bm r}\theta \sigma} 
a_{{\bm r}\theta \sigma}^{\dagger} \left| \psi_0 \right>$. This quantity should be $n_{\theta \sigma}=0.5$ 
if $a_{{\bm r}\theta \sigma}^{\dagger}$ is a well-defined particle. 
The numerically obtained optimized $\theta$ is summarized in Fig. \ref{fig1} (d) for the simplified model, where 
we can see that the optimized $\theta$ varies smoothly with $\lambda$. 
The deviation of $n_{\theta \sigma}$ from 0.5 is found less than 1.2\% for a range of $\lambda$ studied, 
indicating that the ground state is well described by the Kramer's doublet. 

Next, let us study the magnetic properties in region (III) of the phase diagram.  
For this purpose, we first define the structure factor for a local angular momentum operator ${\bm O}^{\mu}_{\bm r}$ 
\begin{eqnarray}
C_{\mu} ({\bm q}) = \frac{1}{N} \sum_{{\bm r},{\bm r}^{\prime}}
e^{i{\bm q} \cdot ({\bm r}-{\bm r}^{\prime})} \left< \psi_0 \right|
{\bm O}_{\bm r}^{\mu} \cdot {\bm O}_{{\bm r}^{\prime}}^{\mu} \left| \psi_0 \right>,
\end{eqnarray}
where $N$ is the number of lattice sites, and $\left| \psi_0 \right>$ indicates the ground state. 
The spin structure factor $C_S ({\bm q})$ is then obtained simply by setting 
\begin{eqnarray}
{\bm O}_{\bm r}^S = {\bm S}_{\bm r}= 
\frac{1}{2} \sum_{\alpha \sigma \sigma^{\prime}} c_{{\bm r} \alpha \sigma}^{\dagger}
{\boldsymbol \sigma}_{\sigma \sigma^{\prime}} c_{{\bm r} \alpha \sigma^{\prime}},
\end{eqnarray}
where ${\boldsymbol \sigma}$ is the vector representation of Pauli matrices. 
The structure factor for the effective total angular momentum 
$C_{J_{\rm eff}} ({\bm q})$ is defined by introducing the following local angular momentum operator 
\begin{eqnarray}
{\bm O}_{\bm r}^{J_{\rm eff}} = {\bm S}_{\bm r} - \sum_{\alpha \alpha^{\prime} \sigma}
c_{{\bm r} \alpha \sigma}^{\dagger} {\bm L}_{\alpha \alpha^{\prime}} c_{{\bm r} \alpha^{\prime} \sigma}, 
\end{eqnarray}
where ${\bm L}$ is the orbital angular momentum operator. 
Note that as long as the local bases are confined within $t_{2g}$ manifold, 
the matrix elements among $t_{2g}$ states are equivalent to those among $p$ states 
apart from minus sign.~\cite{tanabe}
Finally, we define the structure factor for the Kramer's doublet $C_{\rm KD} ({\bm q})$
by introducing the following pseudo-spin 1/2 operator 
\begin{eqnarray}
{\bm O}_{\bm r}^{\rm KD} = \frac{1}{2} \sum_{\sigma \sigma^{\prime}} a_{{\bm r}\theta \sigma}^{\dagger}
{\boldsymbol \sigma}_{\sigma \sigma^{\prime}} a_{{\bm r}\theta \sigma^{\prime}},
\end{eqnarray}
where $a_{{\bm r}\theta \sigma}^{\dagger}$ is given in Eq.~(\ref{eq:kd}). 

The numerical results for these structure factors are shown in Fig. \ref{fig1}. 
For $\lambda=0$, the ground state is the $S=1/2$ Mott insulator, where the hole density 
of $d_{xy}$ orbital is exactly 1 whereby the model is equivalent to 
the single-band Hubbard model. With increasing $\lambda$, 
we can see in  Fig. \ref{fig1} (a)--(c) that the values of $C_{J_{\rm eff}}({\bm q})$ 
approach to those of $C_{S}({\bm q})$ for $\lambda=0$. 
This indicates that the ground state in the limit of large $\lambda$ is represented 
simply by an antiferromagnetic ordering of $J_{\rm eff}=1/2$ angular momentum. 
Finally, we also find in Fig. \ref{fig1} (a)--(c) almost no $\lambda$ dependence on $C_{\rm KD}({\bm q})$, 
which strongly suggests that the ground state for different values of $\lambda$ can be well described by a 
state with alternative alignment of the Kramer's doublet pseudo-spin. 
Therefore, we conclude that the $S=1/2$ and $J_{\rm eff}=1/2$ Mott insulators, 
the two extreme states for $\lambda=0$ and $\infty$, are smoothly connected with no apparent symmetry change. 

\section{\label{sec:vca}Variational Cluster Approximation Study}
We now adopt the VCA method~\cite{vca} based on the SFT~\cite{sft} to study the low-energy one-particle excitations. 
This method takes into account precisely the effects of short-range
static and dynamical correlations, and thus it is superior to a simple mean field approximation. 
The SFT introduces a reference Hamiltonian $H^{\prime}$ with the same two-body interactions
as $H$ but with a different one-body part {${\bm t}^{\prime}$},
and $H^{\prime}$ may be solved numerically exactly on a finite cluster.  An approximate grand potential
for $H$ is given in a functional form by $\Omega ({\bm t}^{\prime}) = \Omega^{\prime} - {\rm Tr} \ln ( - \hat{G}_0^{-1} +
\hat{\Sigma}({\bm t}^{\prime}) ) + {\rm Tr}\ln ( - \hat{G}^{-1}({\bm t}^{\prime}))$,
where $\Omega^{\prime}$, $\hat{\Sigma}({\bm t}^{\prime})$, and $\hat{G} ({\bm t}^{\prime})$ are the 
ground potential, self-energy, and Green's function of the reference system $H^{\prime}$,
respectively.  $\hat{G}_0$ is the non-interacting Green's function of $H$. 
The variational condition $\partial \Omega ({\bm t}^{\prime})/\partial {\bm t}^{\prime} = 0$ determines an
appropriate reference system $H^{\prime}$ which describes the original system $H$ approximately. 

To study the symmetry-broken long-range-ordered states in the VCA, we introduce suitably chosen
fictitious Weiss fields in a set of variational parameters ${\bm t}^{\prime}$.
In this study, we introduce the Weiss field acting on the Kramer's doublet  
\begin{eqnarray}
H_{\rm AF}=h_{\rm AF} \sum_{{\bm r}\sigma} e^{i{\bm Q}\cdot{\bm r}}
a_{{\bm r}\theta \sigma}^{\dagger} a_{{\bm r}\theta \bar{\sigma}}.
\label{eq:weiss}
\end{eqnarray}
with ${\bm Q}=(\pi,\pi)$.  Note that we only consider the Weiss field 
corresponding to the in-plane antiferromagnetic ordering 
since the energy for the states with the out-of-plane antiferromagnetic ordering is always 
higher than that for the in-plane one, as reported previously~\cite{watanabe}. 

\begin{figure}[tb]
\begin{center}
\includegraphics[width=0.7\hsize]{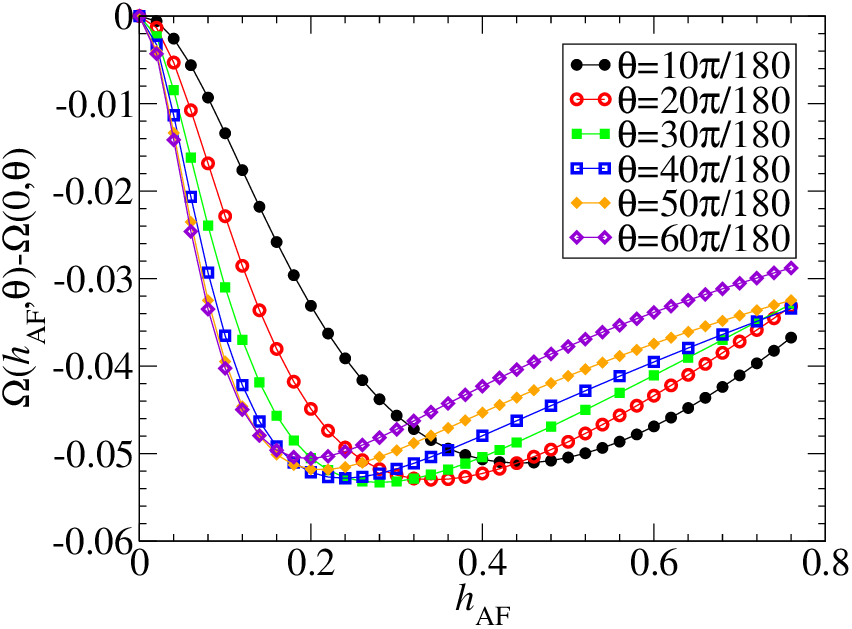}
\caption{(Color online) Variational parameter $h_{\rm AF}$ dependence of the grand potential for various values of 
$\theta$ [Eqs.~(\ref{eq:kd}) and (\ref{eq:weiss})] for $U=1.44$ eV, $J/U=0.15$, and the realistic set of parameters for Sr$_2$IrO$_4$ listed in Table \ref{t1}.  }
\label{fig2}
\end{center}
\end{figure}
Fig. \ref{fig2} shows $\Omega(h_{\rm AF},\theta)-\Omega(0,\theta)$ per site 
for different values of $\theta$ and with $U=1.44$ eV, $J/U = 0.15$, and the realistic set of parameters for Sr$_2$IrO$_4$ 
listed in Table \ref{t1}. 
The fact that this quantity has a minimum at a finite value of $h_{\rm AF}$ indicates that the ground state is 
antiferromagnetically ordered.  Carrying out careful calculations, we also find that the optimal value of $\theta$ is 
$\theta \approx 25^{\circ}$ for this set of model parameters. 

\begin{figure*}[tb]
\begin{center}
\includegraphics[width=0.75\hsize]{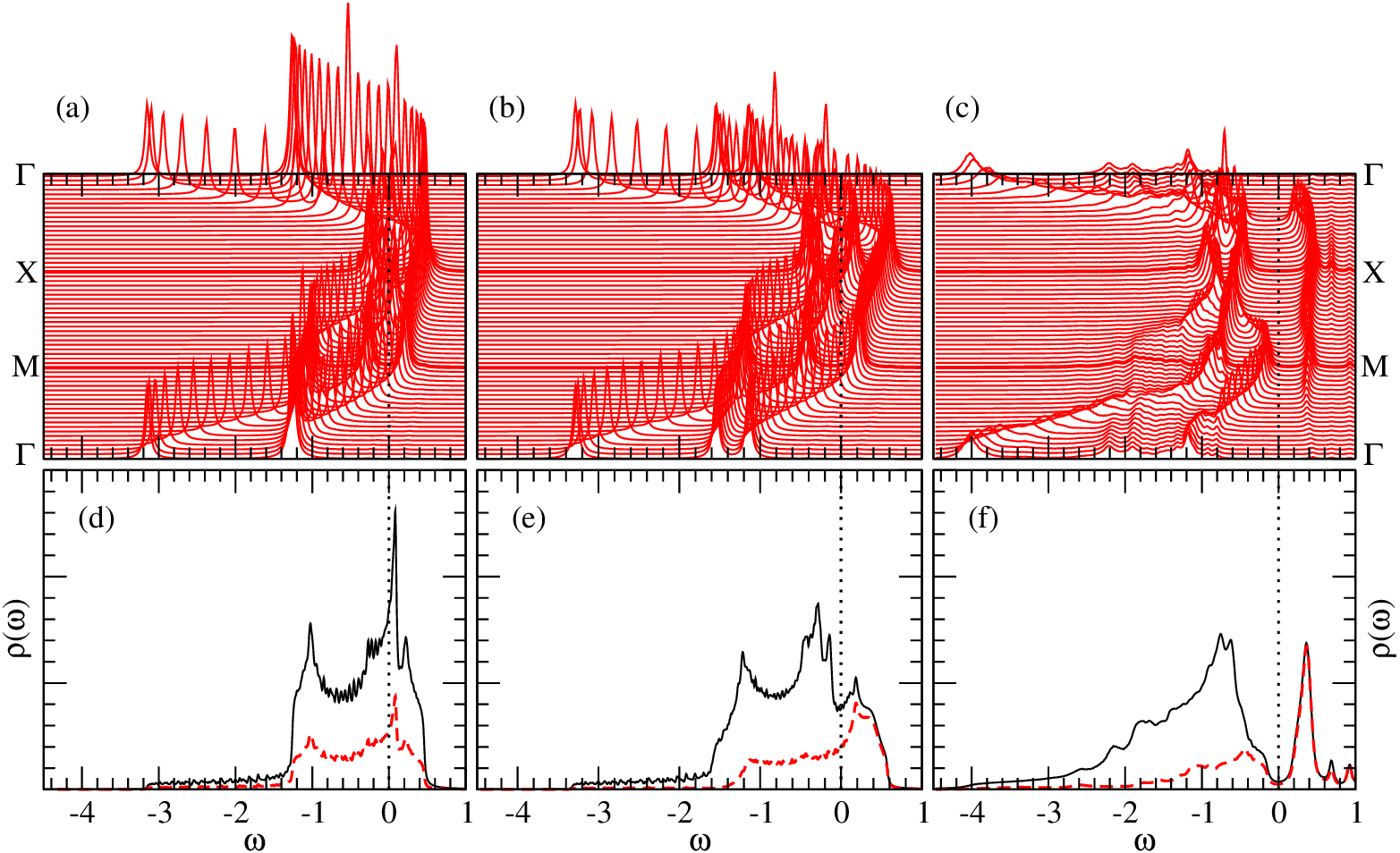}
\caption{(Color online) The momentum resolved one-particle excitation spectra (upper panels) and 
the density of states $\rho(\omega)$ 
(lower panels) for (a) and (d): $\lambda =U =J=0$, (b) and (e): $\lambda = 0.37$ eV and $U=J= 0$, 
and (c) and (f): $\lambda = 0.37$ eV, $U = 1.44$ eV and $J/U=0.15$. 
Other parameters used are the realistic ones for Sr$_2$IrO$_4$ listed in Table \ref{t1}. 
$\Gamma$, M and X correspond to $(k_x, k_y)=(0,0)$, $(\pi,0)$, and $(\pi, \pi)$ in first Brillouin zone, 
respectively. In (d), (e), and (f), black solid (red dashed) lines indicate the total density of states 
(the partial density of states projected onto $J_{\rm eff}=1/2$). 
$\omega = 0$ corresponds to the Fermi level. Delta functions are represented using Lorentzian functions with its 
half-width at half-maximum $\eta = 0.1 t_1$ for (a), (b), and (c), and $\eta = 0.005 t_1$ for (d), (e), and (f). }
\label{fig3}
\end{center}
\end{figure*}

Let us now study the one-particle excitations using the Green's function optimized above. 
The results for the momentum resolved one-particle excitation spectra and the density of states are shown 
in Fig. \ref{fig3}. Here, to understand the effects of $\lambda$ as well as $U$,  we choose three different sets of 
parameters with (a) and (d): $\lambda=U=J=0$, (b) and (e): $\lambda=0.37$ eV and $U=J=0$, 
and (c) and (f):  $\lambda=0.37$ eV, $U=1.44$ eV, and $J/U=0.15$. 

As clearly seen in Fig.~\ref{fig3} (a), when the SOC and the local Coulomb interactions are absent, the bands consist of 
two one-dimension-like narrower bands (ranging from $-1$ eV to 0 eV), which are originated from $d_{yz}$ and $d_{zx}$ 
orbitals, and the remaining band with broader band width extending from $-3.2$ eV to 0.5 eV, which has a characteristic of $d_{xy}$. 
As shown in Fig. \ref{fig3} (d), the projected density of states onto $J_{\rm eff}=1/2$ extends 
to the whole energy band, indicating that $J_{\rm eff}=1/2$ is not a good quantity to 
describe the one-particle excitations. 
When the SOC is turned on in Fig. \ref{fig3} (b) and (e), 
the three different bands are observed, i.e.,  the highest one located from $-1.1$ to 0.6 eV in energy, 
the intermediate one ranging from $-1.4$ to 0.2 eV, and the lowest one extending from $-3.2$ to $-0.2$ eV. 
For this case, we can observe that while the partial density of states projected onto $J_{\rm eff}=1/2$ is still extended to 
the whole energy region,  most of the unoccupied states (from $\sim$0.2 to $\sim$0.6 eV) has a characteristic of 
$J_{\rm eff}=1/2$. Finally, when we include the Coulomb interactions in Fig. \ref{fig3} (c) and (f), 
we can clearly see that the ``upper Hubbard band'', which is located above the Fermi level, is  
well separated from the valence band. We can also see in Fig. \ref{fig3} (f) that 
almost all the unoccupied states is of $J_{\rm eff}=1/2$ characteristic, which provides a numerical evidence that 
the ground state is well characterized by the $J_{\rm eff}=1/2$ 
Mott insulator. This finding is in good qualitative agreement with 
the recent experiments on Sr$_2$IrO$_4$.~\cite{kim1} 


\section{\label{sec:conc}Conclusions}
We have employed the exact diagonalization method to study the ground state phase diagram for 
the three-band Hubbard model with the SOC.  We have found that in the Mott insulating phase the 
ground state can be well described by the Kramer's doublet. This suggests that no apparent 
symmetry change exists between the two extreme states, i.e., the $S=1/2$ Mott insulator and the $J_{\rm eff}=1/2$ Mott insulator, 
which appear for $\lambda=0$ and $\infty$, respectively. 
We have also studied the one-particle excitations using the VCA, and found that both $\lambda$ and 
$U$ are essential to realize the $J_{\rm eff}=1/2$ Mott insulator, where the unoccupied states are mostly of 
$J_{\rm eff}=1/2$ characteristic. 

\section*{Acknowledgment}
The authors thank J. Matsuno and H. Onishi for useful discussions. 
Most of the computation has been done using the RIKEN Cluster of Clusters (RICC) facility.

\end{document}